\documentclass[twocolumn,superscriptaddress,aps,showpacs,floatfix,prl,10pt]{revtex4-2}

\usepackage{graphics}
\usepackage{dcolumn}
\usepackage{bm}
\usepackage{amsmath}
\usepackage{epsfig}
\usepackage{braket}
\usepackage{commath}
\usepackage{float} 
\usepackage[english]{babel}
\usepackage{color}
\usepackage{comment}

\usepackage{color}

\begin{document}

\title{Experimental Higher-Order Interference in a Nonlinear Triple Slit}

\author{Peter Namdar}
\affiliation{University of Vienna, Faculty of Physics, Vienna Center for Quantum Science and Technology (VCQ) and Research Platform for Testing the Quantum and Gravity Interface (TURIS), Boltzmanngasse 5, Vienna A-1090, Austria}

\author{Philipp K. Jenke}
\affiliation{University of Vienna, Faculty of Physics, Vienna Center for Quantum Science and Technology (VCQ) and Research Platform for Testing the Quantum and Gravity Interface (TURIS), Boltzmanngasse 5, Vienna A-1090, Austria}

\author{Irati Alonso Calafell}
\affiliation{University of Vienna, Faculty of Physics, Vienna Center for Quantum Science and Technology (VCQ) and Research Platform for Testing the Quantum and Gravity Interface (TURIS), Boltzmanngasse 5, Vienna A-1090, Austria}

\author{Alessandro Trenti}
\affiliation{University of Vienna, Faculty of Physics, Vienna Center for Quantum Science and Technology (VCQ) and Research Platform for Testing the Quantum and Gravity Interface (TURIS), Boltzmanngasse 5, Vienna A-1090, Austria}
\affiliation{Security and Communication Technologies, Center for Digital Safety and Security, AIT Austrian Institute of Technology GmbH, Giefinggasse 4, 1210 Vienna, Austria}

\author{Milan Radonji\'c}
\affiliation{Department of Physics, Technical University of Kaiserslautern, Kaiserslautern, Germany}
\affiliation{Institute of Physics Belgrade, University of Belgrade, Belgrade, Serbia}

\author{Borivoje Daki\'c}
\affiliation{University of Vienna, Faculty of Physics, Vienna Center for Quantum Science and Technology (VCQ) and Research Platform for Testing the Quantum and Gravity Interface (TURIS), Boltzmanngasse 5, Vienna A-1090, Austria}
\affiliation{Institute for Quantum Optics and Quantum Information (IQOQI), Austrian Academy of Sciences, Boltzmanngasse 3, A-1090 Vienna, Austria}

\author{Philip Walther}
\affiliation{University of Vienna, Faculty of Physics, Vienna Center for Quantum Science and Technology (VCQ) and Research Platform for Testing the Quantum and Gravity Interface (TURIS), Boltzmanngasse 5, Vienna A-1090, Austria}
\affiliation{Christian Doppler Laboratory for Photonic Quantum Computer, Faculty of Physics, University of Vienna, Boltzmanngasse 5, Vienna A-1090, Austria}
\author{Lee A. Rozema}
\affiliation{University of Vienna, Faculty of Physics, Vienna Center for Quantum Science and Technology (VCQ) and Research Platform for Testing the Quantum and Gravity Interface (TURIS), Boltzmanngasse 5, Vienna A-1090, Austria}
\date{\today}

\begin{abstract}
Interference between two waves is a well-known concept in physics, and its generalization to more than two waves is straight-forward.
The order of interference is defined as the number of paths that interfere in a manner that cannot be reduced to patterns of a lower order.
%
%
In practice, second-order interference means that in, say, a triple-slit experiment, the interference pattern when all three slits are open can be predicted from the interference patterns between all possible pairs of slits.
Quantum mechanics is often said to only exhibit second-order interference.
However, this is only true under specific assumptions, typically single-particles undergoing linear evolution.
Here we experimentally show that nonlinear evolution can in fact lead to higher-order interference.
The higher-order interference in our experiment has a simple quantum mechanical description; namely, optical coherent states interacting in a nonlinear medium.
Our work shows that nonlinear evolution could open a loophole for experiments attempting to verify Born's rule by ruling out higher-order interference.
\end{abstract}
\maketitle

Interference between particles is one of the defining phenomena of quantum mechanics, and, perhaps, no scenario exemplifies this better than the double-slit experiment. 
Although the double-slit was originally used to demonstrate the wave nature of light, Feynman later famously said that it ``has in it the heart of quantum mechanics'' \cite{feynman1966feynman}. 
According to standard quantum theory, when one adds additional slits, a measurement of all combinations of double-slit configurations should allow one to predict the final multi-slit interference pattern~\cite{Sorkin}.
More generally, all interference is reducible to double-slit interference.
For this reason, quantum theory is said to exhibit only \textit{second-order} interference.
Sorkin introduced a measurable parameter to determine deviations from this prediction~\cite{Sorkin}, known as the Sorkin-parameter. Finding a non-zero Sorkin-parameter would generally be understood to indicate that our standard formulation of quantum theory, based on Born's rule, is incomplete, or equivalently, that nature requires a description involving higher-order interference.

Experiments finding a zero value for the Sorkin-parameter have already been carried out using a variety of physical systems~\cite{Weihs2010,Weihs2011,Weihs2017,Pleinert2020expt,Laflamme2012, jin2017, Barnea2018, Arndt2017}.
However, in spite of these experiments, it has been shown that under certain circumstances higher-order interference can appear.
In other words, operationally, quantum mechanics exhibits second-order interference only under specific assumptions. If these are violated one can obtain non-zero Sorkin parameter within quantum theory.
To our knowledge, there are currently three mechanisms that are predicted to lead to higher-order interference within quantum theory:
(1) near-field `looped' paths \cite{Yabuki86,RMH2012,Sinha2014,Sinha2015},
(2) multi-particle interference together with coincidence measurements \cite{Pleinert2020theor,horvat2021interference}, 
and (3) nonlinear evolution \cite{rozema2020}.
Note that by nonlinear evolution, we refer to Hamiltonians that are non-quadratic in canonical variables, which is the case for optical nonlinearities or particle-particle interactions.
Under typical conditions these effects are small, and extra effort is required to measure these deviations.
Nonetheless, recent experiments have verified that looped-paths~\cite{Boyd2016,Sinha2018} and multi-particle interference~\cite{Pleinert2020expt} can lead to higher-order interference.

In this letter we present an experiment demonstrating that nonlinear evolution can lead to higher-order interference within quantum theory. 
We achieve this in a `nonlinear triple slit', which is composed of three laser beams interacting in an optically nonlinear crystal.
We further show that, when the nonlinearity is turned off, the higher-order interference disappears.
We stress that the higher-order interference we observe is not a `post-quantum' effect, as discussed in \cite{Sorkin,zyczkowski2008quartic,dakic2014density,lee2017higher}, but it is rather a quantum effect arising in the same sort of apparatus proposed in \cite{Sorkin}, but in a different parameter regime.

In quantum mechanics, every quantum particle can be described by a wave function $\psi$ that is related to the probability of the outcome of a measurement, given by Born's rule as $P(x, t) =  |\psi(x, t)|^2$. 
As a direct consequence of this and the superposition principle, the interference pattern of the double-slit experiment can be described as:
\begin{equation}
\begin{split}
P_{12}(x, t) = |\psi_1(x, t)+\psi_2(x, t)|^2 \\\ = \underbrace{|\psi_1|^2}_{P_1} + \underbrace{|\psi_2|^2}_{P_2} + \underbrace{\psi_1^*\psi_2 + \psi_1\psi_2^*}_{I_{12}},
\label{eq:superposition}
\end{split}
\end{equation}
where $\psi_k(x, t)$ are the single-slit wave functions, and we have dropped the explicit position- and time-dependence in the second line.
In Eq. \ref{eq:superposition}, $P_k$ are the distributions attributed to the single-slits and $I_{12}$ is the interference term.

\begin{figure}
    \centering
    \includegraphics[width=1\linewidth]{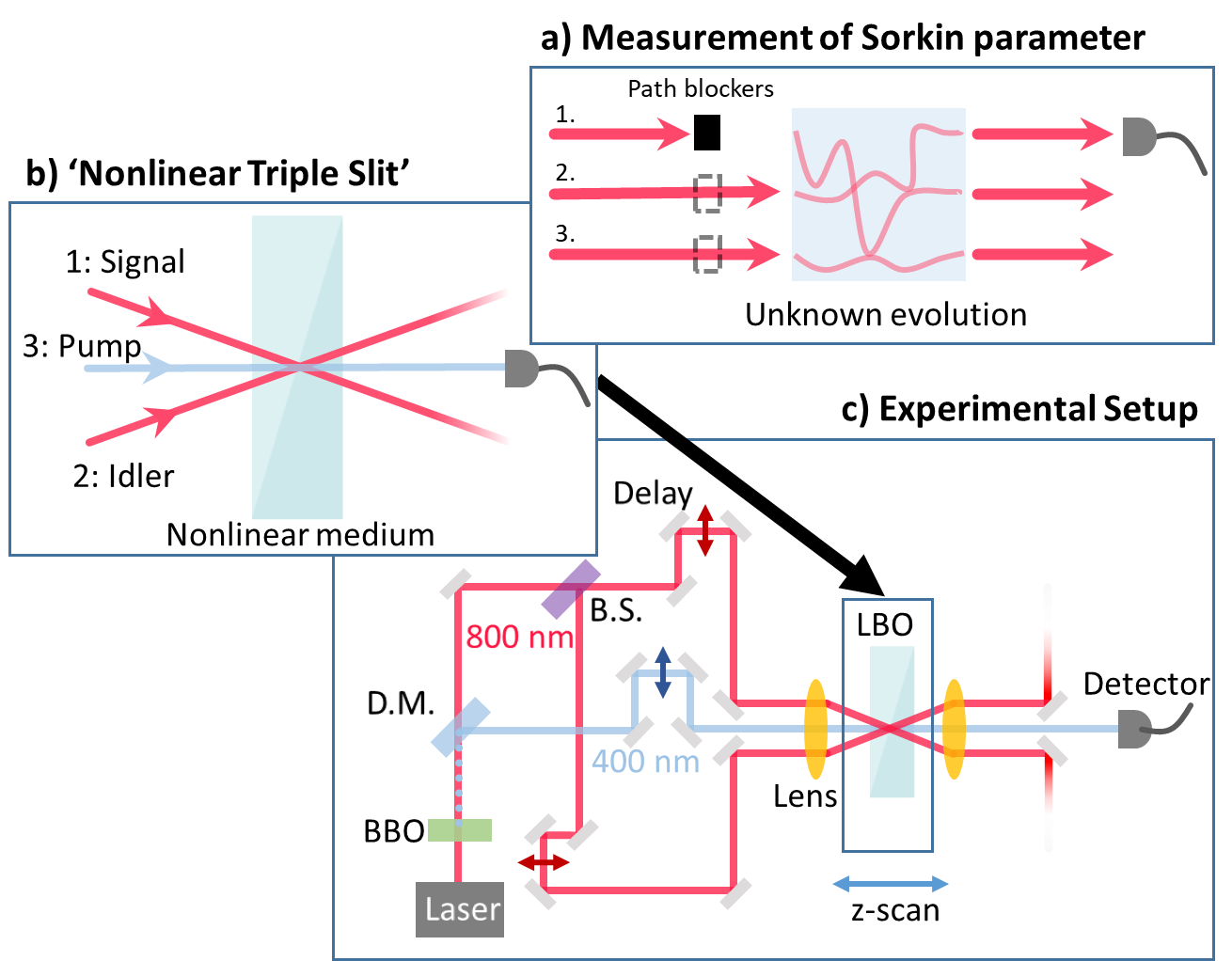}
    \caption{\textbf{(a) Measurement of Sorkin Parameter}: The generic method to measure the Sorkin parameter, on some unknown evolution.  Three (or more) paths are sent through the apparatus and all combinations of paths are blocked and unblocked using the path blockers. A detector placed in a path after the process is used to estimate the probability to detect a particle.  From these measurements, the Sorkin parameter is constructed using Eq. \ref{eq:sorkin_parameter} from the main text.  A measurement of the term $P_{23}$ is pictured.
    \textbf{(b) Nonlinear triple slit}: A pump beam of wavelength $\lambda$ interacts with a signal and an idler beam, both with wavelength $2\lambda$ in a $\chi^{(2)}$-nonlinear crystal.
    \textbf{(c) Experimental setup}: A Ti:Sapphire laser emits light of $\lambda = 800 \mathrm{nm}$ with $\approx140$ fs pulses. The light is partly converted by second-harmonic generation in an beta barium borate (BBO) crystal (3mm length, cut for type-I phase matching)  to $\lambda = 400~\mathrm{nm}$. After being separated by a dichroic mirror, the remaining fundamental beam is split into the signal and idler beams at a 50:50 beamsplitter (B.S.). The length of the beam paths can be adjusted by delay stages to ensure temporal overlap. Shutters are used to block and unblock the individual beams. The three beams are then focused into an Lithium triborat (LBO) crystal (1mm, cut for type-I phase matching). 
    The LBO crystal is mounted on a translation stage to scan it through the focus (z-scan). This simultaneously scans the relative phase between the beams and modulates the strenght of the nonlinear interaction. A photodiode after the crystal is used to measure the optical power in the pump mode.}
    \label{fig:setup}
\end{figure}

The situation is similar in a triple-slit experiment, where the interference pattern can now be described by
\begin{equation}
P_{123} = P_{1} + P_{2} + P_{3} + I_{12} + I_{13} + I_{23},
\label{eq:p123}
\end{equation}
Here, $P_{i}$ are the probabilities to detect a particle with only slit $i$ open, and $I_{ij}$ is the interference term between slits $i$ and $j$, defined in Eq. \ref{eq:superposition}.
Strikingly, there is no third-order interference term; i.e. the interference is reducible to combinations of two-path interference patterns. 
To check for the validity of this formalism, the Sorkin-parameter was introduced, which can be derived directly from Eq. \ref{eq:p123}, by rewriting the interference terms as $I_{jk}=P_{jk}-P_j-P_k$ and moving all terms to the left-hand side of the equation \cite{Weihs2010}:
\begin{equation}
\kappa = P_{123}-P_{12}-P_{23}-P_{13}+P_1+P_2+P_3-P_0,
\label{eq:sorkin_parameter}
\end{equation}
where we have included the term $P_0$--- the probability to observe a detection event with all slits closed---to account for experimental background.
$\kappa$ can be experimentally determined using the apparatus presented in Fig. \ref{fig:setup}a.
For example, $P_{12}$ is the probability to detect a photon (at a given location) when slits 1 and 2 are open.
Assuming linear evolution and single-particle states, standard quantum theory based on Born's rule predicts $\kappa=0$ \cite{Sorkin}.

Following this, the Sorkin-parameter $\kappa$ was experimentally proven to vanish within experimental error in a variety of physical systems by measuring each individual term of Eq. \ref{eq:sorkin_parameter} \cite{Weihs2010,Weihs2011,Kauten2014,Weihs2017,Pleinert2020expt,Laflamme2012, jin2017, Barnea2018, Arndt2017, Boyd2016,Sinha2018}.
Most of these experiments fit the single-particle assumption of Ref.~\cite{Sorkin}.
However, several used multi-particle coherent or thermal states \cite{Weihs2010,Weihs2011,Weihs2017,jin2017,Barnea2018,Arndt2017}, and, nevertheless, found $\kappa=0$.
This is likely because there was no appreciable non-linear evolution involved.
However, detector nonlinearity was noted to be present in Ref. ~\cite{Kauten2014} and was identified as a systematic error leading to $\kappa\neq0$.
Other proposed systems, such as BECs~\cite{leeAtomic}, may be more prone to nonlinear evolution.

Our goal here is to experimentally determine the third-order interference based on nonlinear evolution as introduced in \cite{rozema2020}. 
To understand how nonlinear evolution can lead to higher-order interference, consider three beams (two at frequency $\omega$, labelled signal and idler, and a $2\omega$ pump beam) incident on the nonlinear triple slit presented in Fig. \ref{fig:setup}b.
Although one could measure $\kappa$ with any of the output beams, we will focus on measurements of the pump beam after the nonlinear crystal.
In this case, the single path terms of Eq. \ref{eq:sorkin_parameter} are easy to evaluate: $P_1=P_2=0$, since the detector is mode 3, and in order to measure $P_1$ or $P_2$ light is only sent into mode 1 or 2 respectively.
With light only incident in one mode, there is no mechanism to mix the modes, and generate light in mode 3. Similarly, $P_3=\mathcal{P_\mathrm{pump}} / \mathcal{P_\mathrm{total}}$, which is the input pump power normalized to the total input power, where  $\mathcal{P_\mathrm{total}} = \mathcal{P_\mathrm{signal}} + \mathcal{P_\mathrm{idler}} + \mathcal{P_\mathrm{pump}}$, and $\mathcal{P_\mathrm{signal}}$, $\mathcal{P_\mathrm{idler}}$ and $\mathcal{P_\mathrm{pump}}$, are the powers input into each mode.

The two-path terms correspond to sending light into just two input modes. For example, $P_{12}$ corresponds to sending no light in mode $3$, i.e. blocking the pump.
Nevertheless, light can now be generated in the pump mode via sum-frequency generation (SFG) between signal and idler \cite{Boyd2016}.
Thus $P_{12}=\mathcal{P_\mathrm{SFG}}/\mathcal{P_\mathrm{total}}$, where $\mathcal{P_\mathrm{SFG}}$ is the SFG power generated via the nonlinear mixing in the crystal.
For $P_{13}$ ($P_{23}$), the idler (signal) beam is blocked, and difference-frequency generation (DFG) can occur between the pump and the signal (idler) beams.
We will assume the process is symmetric so $P_{13}=P_{23}=(\mathcal{P_\mathrm{pump}}-\mathcal{P_\mathrm{DFG}})/\mathcal{P_\mathrm{total}}$.

The final term is $P_{123}$, which is the power measured in mode $3$ when all three beams are open.
In this setting both SFG and the two DFG processes will take place, so the three-slit term becomes $P_{123}=(\mathcal{P_\mathrm{pump}}-2\mathcal{P'_\mathrm{DFG}}+\mathcal{P'_\mathrm{SFG}})/\mathcal{P_\mathrm{total}}$. 
Here $\mathcal{P'}$ denotes the fact that the conversion efficiency with all three beams open will be different from the situations with just two beams present.  
This can arise for two reasons.
First, the various nonlinear processes modify the power in the different modes as they propagate through the crystal leading to different net conversion efficiencies.
Second, as we show in the Appendix, when one or two beams are incident, the nonlinear interactions in the crystal are independent of the relative phase. However, the nonlinear interaction between three beams is sensitive to the relative phases, leading to fundamentally different behaviour between the three-path and two-path terms.
Roughly speaking, this phase behaviour arises because when only two beams are incident the relative phase becomes a global phase on the excited nonlinear polarization; on the other hand, the phases do not factor out for three incident beams (see Appendix Eq. \ref{eq:phase}).
With this in mind, we can write the Sorkin parameter as 
\begin{equation}
    \kappa= ((\mathcal{P'_\mathrm{SFG}}-\mathcal{P_\mathrm{SFG}})-2(\mathcal{P'_\mathrm{DFG}}-\mathcal{P_\mathrm{DFG}}))/\mathcal{P_\mathrm{total}}.
\end{equation}
When the nonlinearity is small, for example in the undepleted pump regime,
$\mathcal{P'_\mathrm{SFG}}=\mathcal{P_\mathrm{SFG}}$ and $\mathcal{P'_\mathrm{DFG}}=\mathcal{P_\mathrm{DFG}}$ so $\kappa=0$.
Given the complex nature of the nonlinear interaction, we do not provide explicit forms of DFG and SFG interaction.

In order to maximize the nonlinear interaction in our experiment, we use femtosecond pulsed beams, two of which are depicted as red ($\lambda= 800$ nm) in Fig. \ref{fig:setup} and one is blue ($\lambda= 400$ nm) in Fig. \ref{fig:setup}.  
We call these three beams signal (path 1), idler (path 2) and pump (path 3), which, unless otherwise stated, have powers of approximately $870$ mW, $600$ mW, and $345$ mW, respectively.
As shown in Fig. \ref{fig:setup}c, the three beams, which are generated from a single Ti:sapphire laser, are guided towards the main crystal of our setup and are spatially and temporally overlapped in a 1-mm-thick  Lithium triborat (LBO) crystal  cut  for  type-I  phase-matching.
To satisfy the phase-matching condition, the pump beam is polarized orthogonally to the signal and idler beams.
The beams are focused into the crystal with a $f=25~\mathrm{mm}$ lens, resulting in beam waists of $\approx 30~\mu$m.
Finally, the crystal is mounted on a translation stage which can move it in and out of the focus.
We use this to effectively turn on and off the nonlinear interaction.
When the crystal is not in the focus (the Rayleigh range for our beams is $\approx 3$ mm) then the intensities are too low to generate a measurable nonlinear response.

\begin{figure}
    \centering
    \includegraphics[width=\columnwidth]{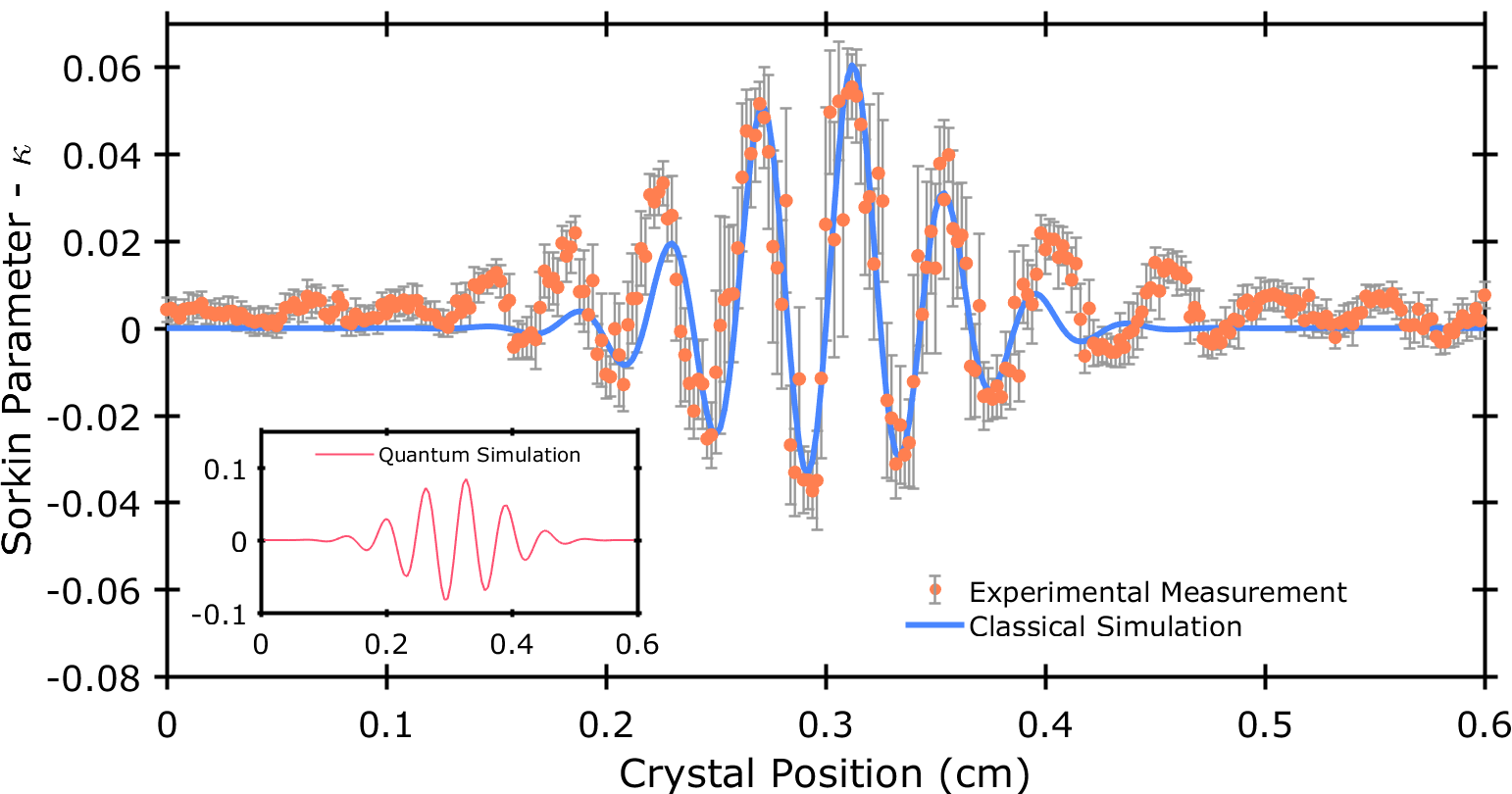}
    \caption{Experimentally measured Sorkin-parameter versus position of the crystal. The orange points are the experimental data. The error bars come from the standard deviation of repeated measurements of the power. The blue curve is the result of a classical simulation of our experiment.
    \textbf{Inset}: A quantum mechanical simulation of the experiment predicts the qualitative features of our data.}
    \label{fig:main_result}
\end{figure}

We will now describe our experimental procedure.
In order to quickly set the individual configurations and measure the corresponding probabilities (as in Fig. \ref{fig:setup}a),
we block various combinations of beams using fast ($\approx 100$ms) optical shutters.
The measurements are performed with a standard optical power meter, measuring the pump beam after the crystal.
For each power measurement, we in fact measure the power 500 times, and average the result.
A single cycle of the experiment, measuring the power for each of the 8 configurations (i.e. the individual terms of Eq. \ref{eq:sorkin_parameter}), takes $\approx 1$ minute.
To vary the strength of the nonlinear interaction we then translate the crystal from a position behind the common focal point of the beams, through the focus, and then out of focus in front of the beams as in Ref. \cite{calafell2021giant}.
In absolute numbers, this corresponds to moving the crystal from $z = 0.00~\mathrm{cm}$ to $z = 0.60~\mathrm{cm}$, with the focus at  $z \approx 0.35~\mathrm{cm}$, the result of this measurement is plotted in Fig. \ref{fig:main_result}.

In addition to varying the strength of the nonlinear interaction, this procedure induces a relative phase between the beams (as we describe in detail in the Appendix).
This phase only affects the $P_{123}$ term, and is the origin of the fringes in Fig. \ref{fig:main_result}.
This, again, shows that the higher-order interference pattern with all three slits open is qualitatively different from the second-order interference patterns.
Given the phase dependence of this term, we perform additional stability measurements (Fig. \ref{fig:stability} of the Appendix), finding that our setup is passively phase stable for over 90 minutes.

To model our experiment we use a classical and a quantum method, both of which are detailed in the Appendix. For the quantum approach, we start with the Hamiltonian
\begin{equation}
\hat{H}=\hbar\omega \hat{a}^\dagger_1\hat{a}_1 + \hbar\omega \hat{a}^\dagger_2\hat{a}_2 + 2\hbar\omega \hat{a}^\dagger_3\hat{a}_3 + i\hbar\chi^{(2)} (\hat{a}_1\hat{a}_2\hat{a}_3^\dagger - \hat{a}_1^\dagger\hat{a}_2^\dagger\hat{a}_3)
\end{equation}
and we assume coherent states $\ket{\alpha_1}_\omega, \ket{\alpha_2}_\omega$ and $\ket{\alpha_3}_{2\omega}$ are input into the various modes.
We then apply 6th order perturbation theory \cite{Sakurai}, and compute the average photon number in mode 3 $\langle\hat{n_3}\rangle$ after the interaction.
As discussed in the Appendix, we use our experimentally measured conversion efficiencies to set the product of the nonlinearity $\chi^{(2)}$ and the interaction time.
The final result of this calculation is plotted in the inset of Fig. 2.
This clearly shows a non-zero value of $\kappa$, demonstrating the presence of higher-order interference described by quantum theory.

\begin{figure}
    \centering
    \includegraphics[width=\linewidth]{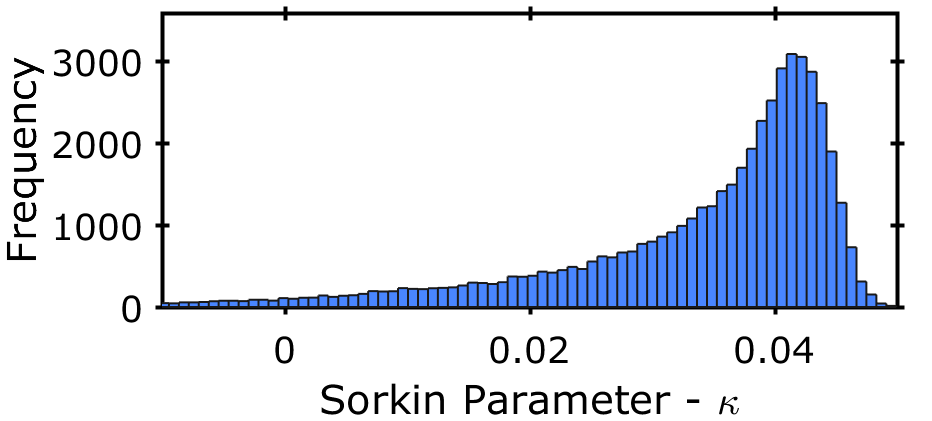}
    \caption{A histogram of 50000 repeated measurements of the Sorkin parameter at a fixed crystal position.
    For these data, the crystal was set to a position corresponding to a maximal value of $\kappa$. The asymmetric distribution occurs because phase fluctuations can only reduce this value.
    The net measurement time for these data was approximately one hour.
    }
    \label{fig:sorkin_static}
\end{figure}

The qualitative features of our experimental data, shown in the main panel of Fig. \ref{fig:main_result}, are reproduced by the quantum model.
To fit to the data, we use a classical model as this allows us to more readily include the different efficiencies for the different processes. In our classical model, we numerically solve the coupled wave equations governing the interaction of three beams in a nonlinear crystal \cite{Boyd2016}.
In more detail, assuming collinear propagation, perfect phase-matching $\Delta k = k_1 + k_2 - k_3=0$, and degenerate signal and idler wavelengths $\omega_\mathrm{1} = \omega_\mathrm{2} = \frac{1}{2}\omega_\mathrm{3}$,
 we can write the coupled-wave equations as:
\begin{eqnarray}
\label{eq:scde1}
\frac{dE_{1}}{dz}&=&i g E_3 E_2^*\\
\label{eq:scde2}
\frac{dE_{2}}{dz}&=&i g E_3 E_1^*\\
\label{eq:scde3}
\frac{dE_{3}}{dz}&=&i 2 g E_1 E_2,
\end{eqnarray}\\
where $g=\frac{2 d_\mathrm{eff}\omega_1}{c}$ and $d_\mathrm{eff}$ is the nonlinear strength of the crystal.
These equations can be solved numerically; we do so using MATLAB. Our code solving these equations is available at \cite{hoiDATA}.
In the Appendix we discuss how we fit this model to our experimental data by measuring the DFG and SFG output powers for each pair of beams and using different coupling constants $g$ for each differential equation from Eqs. \ref{eq:scde1}-\ref{eq:scde3}.
After fixing these parameters, the only remaining free parameter is the angle at which the beams intersect in the crystal, which sets the fringe period.
The nominal angle of intersection for our crystal is $3^\circ$ (given by the phase-matching conditions), but we find a better fit to our data with $6.2^\circ$.
This deviation likely comes from the large numerical aperture of our focusing lens.
The resulting simulation is plotted as the solid line in Fig. \ref{fig:main_result}.  We observe good agreement between our experimentally measured Sorkin parameter and that simulated by our model.
Both our quantum and classical theoretical models, together with our experimental data, clearly show the presence of higher-order interference, without requiring any exotic physics.

To better quantify this effect, we performed a longer measurement with the crystal set to the point at which we observed a maximum $\kappa$ ($Z\approx0.31$ cm).
We again measured the individual terms of Eq. \ref{eq:sorkin_parameter} by blocking and unblocking each of the beam paths $100$ times and measuring the power. Since for each power measurement, we measured the power $500$ times, this results in $50,000$ individual measurements of $\kappa$ in $\approx 1$ hour.
A histogram of the observed values is shown in Fig. \ref{fig:sorkin_static}.
The asymmetric distribution is due to phase fluctuations which occurred during the long measurement time.  
Since the experiment was aligned at a maximum of the fringe, fluctuations predominately decrease $\kappa$.
To obtain our final value, we do not use the data presented in Fig. \ref{fig:sorkin_static}, since this always uses the first measured values of power to compute a value for $\kappa$, and there is no \textit{a priori} reason to make this assumption.
Instead, we estimate each term of Eq. \ref{eq:sorkin_parameter} directly by taking the mean of the data presented in the Appendix Fig. \ref{fig:individual_histograms}. We also compute the standard error from the standard deviation of these data, and arrive at $\kappa=0.0334\pm0.0002$, where the final error bar is obtained by standard propagation of error through Eq. \ref{eq:sorkin_parameter}.

To conclude, in this letter we have presented an experiment wherein we can turn on and off higher-order interference by modulating the nonlinear interactions in our system.
Together with Ref. \cite{rozema2020}, this demonstrates that multi-particle input states on their own are not sufficient to observe higher-order interference, rather nonlinearity is the key.
Since many experiments searching for violations of Born's rule have used multi-particle states, our work shows that such tests must also consider sources of nonlinearities before claiming any deviation from quantum theory.
Finally, we stress again that our work does not imply a failing of Born's rule.
One could, of course, still apply Born's rule to our experiment.
To do so one needs to compute the quantum state \textit{after} the nonlinear interaction, which cannot be constructed by taking a linear superposition of the states in the input modes.

The authors acknowledge support from the research platform TURIS, the  Austrian Science Fund (FWF) through BeyondC (F7113),  Reseach Group 5 (FG5), and TAI 483-N, and from the AFOSR via PhoQuGraph (FA8655-20-1-7030). 
B.D. acknowledges support from the Austrian Science Fund (FWF) through BeyondC (F7112). 
A.T. acknowledges support from the European Union’s Horizon 2020 research and innovation program under the Marie Skłodowska-Curie grant agreement no. 801110 and the Austrian Federal Ministry of Education, Science and Research (BMBWF).
M. R. acknowledges financial support by the Deutsche Forschungsgemeinschaft (DFG, German Research Foundation) via the Collaborative Research Center SFB/TR185 (Project No. 277625399) and via the Research Unit FOR 2247 (Project No. PE 530/6-1).

\bibliography{HOI_in_a_NL_triple_slit.bbl}

\begin{thebibliography}{29}%
\makeatletter
\providecommand \@ifxundefined [1]{%
 \@ifx{#1\undefined}
}%
\providecommand \@ifnum [1]{%
 \ifnum #1\expandafter \@firstoftwo
 \else \expandafter \@secondoftwo
 \fi
}%
\providecommand \@ifx [1]{%
 \ifx #1\expandafter \@firstoftwo
 \else \expandafter \@secondoftwo
 \fi
}%
\providecommand \natexlab [1]{#1}%
\providecommand \enquote  [1]{``#1''}%
\providecommand \bibnamefont  [1]{#1}%
\providecommand \bibfnamefont [1]{#1}%
\providecommand \citenamefont [1]{#1}%
\providecommand \href@noop [0]{\@secondoftwo}%
\providecommand \href [0]{\begingroup \@sanitize@url \@href}%
\providecommand \@href[1]{\@@startlink{#1}\@@href}%
\providecommand \@@href[1]{\endgroup#1\@@endlink}%
\providecommand \@sanitize@url [0]{\catcode `\\12\catcode `\$12\catcode
  `\&12\catcode `\#12\catcode `\^12\catcode `\_12\catcode `\%12\relax}%
\providecommand \@@startlink[1]{}%
\providecommand \@@endlink[0]{}%
\providecommand \url  [0]{\begingroup\@sanitize@url \@url }%
\providecommand \@url [1]{\endgroup\@href {#1}{\urlprefix }}%
\providecommand \urlprefix  [0]{URL }%
\providecommand \Eprint [0]{\href }%
\providecommand \doibase [0]{http://dx.doi.org/}%
\providecommand \selectlanguage [0]{\@gobble}%
\providecommand \bibinfo  [0]{\@secondoftwo}%
\providecommand \bibfield  [0]{\@secondoftwo}%
\providecommand \translation [1]{[#1]}%
\providecommand \BibitemOpen [0]{}%
\providecommand \bibitemStop [0]{}%
\providecommand \bibitemNoStop [0]{.\EOS\space}%
\providecommand \EOS [0]{\spacefactor3000\relax}%
\providecommand \BibitemShut  [1]{\csname bibitem#1\endcsname}%
\let\auto@bib@innerbib\@empty
\bibitem [{\citenamefont {Feynman}\ \emph {et~al.}(1966)\citenamefont
  {Feynman}, \citenamefont {Leighton},\ and\ \citenamefont
  {Sands}}]{feynman1966feynman}%
  \BibitemOpen
  \bibfield  {author} {\bibinfo {author} {\bibfnamefont {R.P.}\ \bibnamefont
  {Feynman}}, \bibinfo {author} {\bibfnamefont {R.B.}\ \bibnamefont
  {Leighton}}, \ and\ \bibinfo {author} {\bibfnamefont {M.L.}\ \bibnamefont
  {Sands}},\ }\href {https://books.google.at/books?id=6f3OzAEACAAJ} {\emph
  {\bibinfo {title} {The Feynman Lectures on Physics: Quantum mechanics}}},\
  The Feynman Lectures on Physics\ (\bibinfo  {publisher} {Pearson India
  Education Services Pvt Limited},\ \bibinfo {year} {1966})\BibitemShut
  {NoStop}%
\bibitem [{\citenamefont {Sorkin}(1994)}]{Sorkin}%
  \BibitemOpen
  \bibfield  {author} {\bibinfo {author} {\bibfnamefont {R.~D.}\ \bibnamefont
  {Sorkin}},\ }\bibfield  {title} {\enquote {\bibinfo {title} {Quantum
  mechanics as quantum measure theory},}\ }\href@noop {} {\bibfield  {journal}
  {\bibinfo  {journal} {Mod. Phys. Lett. A}\ }\textbf {\bibinfo {volume} {9}},\
  \bibinfo {pages} {3119} (\bibinfo {year} {1994})}\BibitemShut {NoStop}%
\bibitem [{\citenamefont {Sinha}\ \emph {et~al.}(2010)\citenamefont {Sinha},
  \citenamefont {Couteau}, \citenamefont {Jennewein}, \citenamefont
  {Laflamme},\ and\ \citenamefont {Weihs}}]{Weihs2010}%
  \BibitemOpen
  \bibfield  {author} {\bibinfo {author} {\bibfnamefont {U.}~\bibnamefont
  {Sinha}}, \bibinfo {author} {\bibfnamefont {C.}~\bibnamefont {Couteau}},
  \bibinfo {author} {\bibfnamefont {T.}~\bibnamefont {Jennewein}}, \bibinfo
  {author} {\bibfnamefont {R.}~\bibnamefont {Laflamme}}, \ and\ \bibinfo
  {author} {\bibfnamefont {G.}~\bibnamefont {Weihs}},\ }\bibfield  {title}
  {\enquote {\bibinfo {title} {Ruling out multi-order interference in quantum
  mechanics},}\ }\href@noop {} {\bibfield  {journal} {\bibinfo  {journal}
  {Science}\ }\textbf {\bibinfo {volume} {329}},\ \bibinfo {pages} {418}
  (\bibinfo {year} {2010})}\BibitemShut {NoStop}%
\bibitem [{\citenamefont {S\"ollner}\ \emph {et~al.}(2011)\citenamefont
  {S\"ollner}, \citenamefont {Gsch\"osser}, \citenamefont {Mai}, \citenamefont
  {Pressl}, \citenamefont {V\"or\"os},\ and\ \citenamefont
  {Weihs}}]{Weihs2011}%
  \BibitemOpen
  \bibfield  {author} {\bibinfo {author} {\bibfnamefont {I.}~\bibnamefont
  {S\"ollner}}, \bibinfo {author} {\bibfnamefont {B.}~\bibnamefont
  {Gsch\"osser}}, \bibinfo {author} {\bibfnamefont {P.}~\bibnamefont {Mai}},
  \bibinfo {author} {\bibfnamefont {B.}~\bibnamefont {Pressl}}, \bibinfo
  {author} {\bibfnamefont {Z.}~\bibnamefont {V\"or\"os}}, \ and\ \bibinfo
  {author} {\bibfnamefont {G.}~\bibnamefont {Weihs}},\ }\bibfield  {title}
  {\enquote {\bibinfo {title} {Testing born's rule in quantum mechanics for
  three mutually exclusive events},}\ }\href@noop {} {\bibfield  {journal}
  {\bibinfo  {journal} {Found. Phys.}\ }\textbf {\bibinfo {volume} {42}},\
  \bibinfo {pages} {742} (\bibinfo {year} {2011})}\BibitemShut {NoStop}%
\bibitem [{\citenamefont {Kauten}\ \emph {et~al.}(2017)\citenamefont {Kauten},
  \citenamefont {Keil}, \citenamefont {Kaufmann}, \citenamefont {Pressl},
  \citenamefont {Brukner},\ and\ \citenamefont {Weihs}}]{Weihs2017}%
  \BibitemOpen
  \bibfield  {author} {\bibinfo {author} {\bibfnamefont {T.}~\bibnamefont
  {Kauten}}, \bibinfo {author} {\bibfnamefont {R.}~\bibnamefont {Keil}},
  \bibinfo {author} {\bibfnamefont {T.}~\bibnamefont {Kaufmann}}, \bibinfo
  {author} {\bibfnamefont {B.}~\bibnamefont {Pressl}}, \bibinfo {author}
  {\bibfnamefont {{\v C}.}~\bibnamefont {Brukner}}, \ and\ \bibinfo {author}
  {\bibfnamefont {G.}~\bibnamefont {Weihs}},\ }\bibfield  {title} {\enquote
  {\bibinfo {title} {Obtaining tight bounds on higher-order interferences with
  a 5-path interferometer},}\ }\href@noop {} {\bibfield  {journal} {\bibinfo
  {journal} {New J. Phys.}\ }\textbf {\bibinfo {volume} {19}},\ \bibinfo
  {pages} {033017} (\bibinfo {year} {2017})}\BibitemShut {NoStop}%
\bibitem [{\citenamefont {Pleinert}\ \emph {et~al.}(2021)\citenamefont
  {Pleinert}, \citenamefont {Rueda}, \citenamefont {Lutz},\ and\ \citenamefont
  {von Zanthier}}]{Pleinert2020expt}%
  \BibitemOpen
  \bibfield  {author} {\bibinfo {author} {\bibfnamefont {Marc-Oliver}\
  \bibnamefont {Pleinert}}, \bibinfo {author} {\bibfnamefont {Alfredo}\
  \bibnamefont {Rueda}}, \bibinfo {author} {\bibfnamefont {Eric}\ \bibnamefont
  {Lutz}}, \ and\ \bibinfo {author} {\bibfnamefont {Joachim}\ \bibnamefont {von
  Zanthier}},\ }\bibfield  {title} {\enquote {\bibinfo {title} {Testing
  higher-order quantum interference with many-particle states},}\ }\href@noop
  {} {\bibfield  {journal} {\bibinfo  {journal} {Physical Review Letters}\
  }\textbf {\bibinfo {volume} {126}},\ \bibinfo {pages} {190401} (\bibinfo
  {year} {2021})}\BibitemShut {NoStop}%
\bibitem [{\citenamefont {Park}\ \emph {et~al.}(2012)\citenamefont {Park},
  \citenamefont {Moussa},\ and\ \citenamefont {Laflamme}}]{Laflamme2012}%
  \BibitemOpen
  \bibfield  {author} {\bibinfo {author} {\bibfnamefont {D.}~\bibnamefont
  {Park}}, \bibinfo {author} {\bibfnamefont {O.}~\bibnamefont {Moussa}}, \ and\
  \bibinfo {author} {\bibfnamefont {R.}~\bibnamefont {Laflamme}},\ }\bibfield
  {title} {\enquote {\bibinfo {title} {Three path interference using nuclear
  magnetic resonance: A test of the consistency of born's rule},}\ }\href@noop
  {} {\bibfield  {journal} {\bibinfo  {journal} {New J. Phys.}\ }\textbf
  {\bibinfo {volume} {14}},\ \bibinfo {pages} {113025} (\bibinfo {year}
  {2012})}\BibitemShut {NoStop}%
\bibitem [{\citenamefont {Jin}\ \emph {et~al.}(2017)\citenamefont {Jin},
  \citenamefont {Liu}, \citenamefont {Geng}, \citenamefont {Huang},
  \citenamefont {Ma}, \citenamefont {Shi}, \citenamefont {Duan}, \citenamefont
  {Shi}, \citenamefont {Rong},\ and\ \citenamefont {Du}}]{jin2017}%
  \BibitemOpen
  \bibfield  {author} {\bibinfo {author} {\bibfnamefont {Fangzhou}\
  \bibnamefont {Jin}}, \bibinfo {author} {\bibfnamefont {Ying}\ \bibnamefont
  {Liu}}, \bibinfo {author} {\bibfnamefont {Jianpei}\ \bibnamefont {Geng}},
  \bibinfo {author} {\bibfnamefont {Pu}~\bibnamefont {Huang}}, \bibinfo
  {author} {\bibfnamefont {Wenchao}\ \bibnamefont {Ma}}, \bibinfo {author}
  {\bibfnamefont {Mingjun}\ \bibnamefont {Shi}}, \bibinfo {author}
  {\bibfnamefont {Chang-Kui}\ \bibnamefont {Duan}}, \bibinfo {author}
  {\bibfnamefont {Fazhan}\ \bibnamefont {Shi}}, \bibinfo {author}
  {\bibfnamefont {Xing}\ \bibnamefont {Rong}}, \ and\ \bibinfo {author}
  {\bibfnamefont {Jiangfeng}\ \bibnamefont {Du}},\ }\bibfield  {title}
  {\enquote {\bibinfo {title} {Experimental test of born's rule by inspecting
  third-order quantum interference on a single spin in solids},}\ }\href@noop
  {} {\bibfield  {journal} {\bibinfo  {journal} {Physical Review A}\ }\textbf
  {\bibinfo {volume} {95}},\ \bibinfo {pages} {012107} (\bibinfo {year}
  {2017})}\BibitemShut {NoStop}%
\bibitem [{\citenamefont {Barnea}\ \emph {et~al.}(2018)\citenamefont {Barnea},
  \citenamefont {Cheshnovsky},\ and\ \citenamefont {Even}}]{Barnea2018}%
  \BibitemOpen
  \bibfield  {author} {\bibinfo {author} {\bibfnamefont {A.~Ronny}\
  \bibnamefont {Barnea}}, \bibinfo {author} {\bibfnamefont {Ori}\ \bibnamefont
  {Cheshnovsky}}, \ and\ \bibinfo {author} {\bibfnamefont {Uzi}\ \bibnamefont
  {Even}},\ }\bibfield  {title} {\enquote {\bibinfo {title} {Matter-wave
  diffraction approaching limits predicted by feynman path integrals for
  multipath interference},}\ }\href {\doibase 10.1103/PhysRevA.97.023601}
  {\bibfield  {journal} {\bibinfo  {journal} {Phys. Rev. A}\ }\textbf {\bibinfo
  {volume} {97}},\ \bibinfo {pages} {023601} (\bibinfo {year}
  {2018})}\BibitemShut {NoStop}%
\bibitem [{\citenamefont {Cotter}\ \emph {et~al.}(2017)\citenamefont {Cotter},
  \citenamefont {Brand}, \citenamefont {Knobloch}, \citenamefont {Lilach},
  \citenamefont {Cheshnovsky},\ and\ \citenamefont {Arndt}}]{Arndt2017}%
  \BibitemOpen
  \bibfield  {author} {\bibinfo {author} {\bibfnamefont {J.~P.}\ \bibnamefont
  {Cotter}}, \bibinfo {author} {\bibfnamefont {C.}~\bibnamefont {Brand}},
  \bibinfo {author} {\bibfnamefont {C.}~\bibnamefont {Knobloch}}, \bibinfo
  {author} {\bibfnamefont {Y.}~\bibnamefont {Lilach}}, \bibinfo {author}
  {\bibfnamefont {O.}~\bibnamefont {Cheshnovsky}}, \ and\ \bibinfo {author}
  {\bibfnamefont {M.}~\bibnamefont {Arndt}},\ }\href@noop {} {\bibfield
  {journal} {\bibinfo  {journal} {Sci. Adv.}\ }\textbf {\bibinfo {volume}
  {3}},\ \bibinfo {pages} {e1602478} (\bibinfo {year} {2017})}\BibitemShut
  {NoStop}%
\bibitem [{\citenamefont {Yabuki}(1986)}]{Yabuki86}%
  \BibitemOpen
  \bibfield  {author} {\bibinfo {author} {\bibfnamefont {H.}~\bibnamefont
  {Yabuki}},\ }\bibfield  {title} {\enquote {\bibinfo {title} {Feynman path
  integrals in the young double-slit experiment},}\ }\href@noop {} {\bibfield
  {journal} {\bibinfo  {journal} {Int. J. Theor. Phys.}\ }\textbf {\bibinfo
  {volume} {25}},\ \bibinfo {pages} {159} (\bibinfo {year} {1986})}\BibitemShut
  {NoStop}%
\bibitem [{\citenamefont {Raedt}\ \emph {et~al.}(2012)\citenamefont {Raedt},
  \citenamefont {Michielsen},\ and\ \citenamefont {Hess}}]{RMH2012}%
  \BibitemOpen
  \bibfield  {author} {\bibinfo {author} {\bibfnamefont {H.~De}\ \bibnamefont
  {Raedt}}, \bibinfo {author} {\bibfnamefont {K.}~\bibnamefont {Michielsen}}, \
  and\ \bibinfo {author} {\bibfnamefont {K.}~\bibnamefont {Hess}},\ }\bibfield
  {title} {\enquote {\bibinfo {title} {Analysis of multipath interference in
  three-slit experiments},}\ }\href@noop {} {\bibfield  {journal} {\bibinfo
  {journal} {Phys. Rev. A}\ }\textbf {\bibinfo {volume} {85}},\ \bibinfo
  {pages} {012101} (\bibinfo {year} {2012})}\BibitemShut {NoStop}%
\bibitem [{\citenamefont {Sawant}\ \emph {et~al.}(2014)\citenamefont {Sawant},
  \citenamefont {Samuel}, \citenamefont {Sinha}, \citenamefont {Sinha},\ and\
  \citenamefont {Sinha}}]{Sinha2014}%
  \BibitemOpen
  \bibfield  {author} {\bibinfo {author} {\bibfnamefont {R.}~\bibnamefont
  {Sawant}}, \bibinfo {author} {\bibfnamefont {J.}~\bibnamefont {Samuel}},
  \bibinfo {author} {\bibfnamefont {A.}~\bibnamefont {Sinha}}, \bibinfo
  {author} {\bibfnamefont {S.}~\bibnamefont {Sinha}}, \ and\ \bibinfo {author}
  {\bibfnamefont {U.}~\bibnamefont {Sinha}},\ }\bibfield  {title} {\enquote
  {\bibinfo {title} {Nonclassical paths in quantum interference experiments},}\
  }\href@noop {} {\bibfield  {journal} {\bibinfo  {journal} {Phys. Rev. Lett.}\
  }\textbf {\bibinfo {volume} {113}},\ \bibinfo {pages} {120406} (\bibinfo
  {year} {2014})}\BibitemShut {NoStop}%
\bibitem [{\citenamefont {Sinha}\ \emph {et~al.}(2015)\citenamefont {Sinha},
  \citenamefont {Vijay},\ and\ \citenamefont {Sinha}}]{Sinha2015}%
  \BibitemOpen
  \bibfield  {author} {\bibinfo {author} {\bibfnamefont {A.}~\bibnamefont
  {Sinha}}, \bibinfo {author} {\bibfnamefont {A.~H.}\ \bibnamefont {Vijay}}, \
  and\ \bibinfo {author} {\bibfnamefont {U.}~\bibnamefont {Sinha}},\ }\bibfield
   {title} {\enquote {\bibinfo {title} {On the superposition principle in
  interference experiments},}\ }\href@noop {} {\bibfield  {journal} {\bibinfo
  {journal} {Sci. Rep.}\ }\textbf {\bibinfo {volume} {5}},\ \bibinfo {pages}
  {10304} (\bibinfo {year} {2015})}\BibitemShut {NoStop}%
\bibitem [{\citenamefont {Pleinert}\ \emph {et~al.}(2020)\citenamefont
  {Pleinert}, \citenamefont {von Zanthier},\ and\ \citenamefont
  {Lutz}}]{Pleinert2020theor}%
  \BibitemOpen
  \bibfield  {author} {\bibinfo {author} {\bibfnamefont {Marc-Oliver}\
  \bibnamefont {Pleinert}}, \bibinfo {author} {\bibfnamefont {Joachim}\
  \bibnamefont {von Zanthier}}, \ and\ \bibinfo {author} {\bibfnamefont {Eric}\
  \bibnamefont {Lutz}},\ }\bibfield  {title} {\enquote {\bibinfo {title}
  {Many-particle interference to test born's rule},}\ }\href {\doibase
  10.1103/PhysRevResearch.2.012051} {\bibfield  {journal} {\bibinfo  {journal}
  {Phys. Rev. Research}\ }\textbf {\bibinfo {volume} {2}},\ \bibinfo {pages}
  {012051} (\bibinfo {year} {2020})}\BibitemShut {NoStop}%
\bibitem [{\citenamefont {Horvat}\ and\ \citenamefont
  {Daki{\'c}}(2021)}]{horvat2021interference}%
  \BibitemOpen
  \bibfield  {author} {\bibinfo {author} {\bibfnamefont {Sebastian}\
  \bibnamefont {Horvat}}\ and\ \bibinfo {author} {\bibfnamefont {Borivoje}\
  \bibnamefont {Daki{\'c}}},\ }\bibfield  {title} {\enquote {\bibinfo {title}
  {Interference as an information-theoretic game},}\ }\href@noop {} {\bibfield
  {journal} {\bibinfo  {journal} {Quantum}\ }\textbf {\bibinfo {volume} {5}},\
  \bibinfo {pages} {404} (\bibinfo {year} {2021})}\BibitemShut {NoStop}%
\bibitem [{\citenamefont {Rozema}\ \emph {et~al.}(2021)\citenamefont {Rozema},
  \citenamefont {Zhuo}, \citenamefont {Paterek},\ and\ \citenamefont
  {Daki\ifmmode~\acute{c}\else \'{c}\fi{}}}]{rozema2020}%
  \BibitemOpen
  \bibfield  {author} {\bibinfo {author} {\bibfnamefont {Lee~A.}\ \bibnamefont
  {Rozema}}, \bibinfo {author} {\bibfnamefont {Zhao}\ \bibnamefont {Zhuo}},
  \bibinfo {author} {\bibfnamefont {Tomasz}\ \bibnamefont {Paterek}}, \ and\
  \bibinfo {author} {\bibfnamefont {Borivoje}\ \bibnamefont
  {Daki\ifmmode~\acute{c}\else \'{c}\fi{}}},\ }\bibfield  {title} {\enquote
  {\bibinfo {title} {Higher-order interference between multiple quantum
  particles interacting nonlinearly},}\ }\href {\doibase
  10.1103/PhysRevA.103.052204} {\bibfield  {journal} {\bibinfo  {journal}
  {Phys. Rev. A}\ }\textbf {\bibinfo {volume} {103}},\ \bibinfo {pages}
  {052204} (\bibinfo {year} {2021})}\BibitemShut {NoStop}%
\bibitem [{\citenamefont {Magana-Loaiza}\ \emph {et~al.}(2016)\citenamefont
  {Magana-Loaiza}, \citenamefont {Leon}, \citenamefont {Mirhosseini},
  \citenamefont {Fickler}, \citenamefont {Safari}, \citenamefont {Mick},
  \citenamefont {McIntyre}, \citenamefont {Banzer}, \citenamefont {Rodenburg},
  \citenamefont {Leuchs},\ and\ \citenamefont {Boyd}}]{Boyd2016}%
  \BibitemOpen
  \bibfield  {author} {\bibinfo {author} {\bibfnamefont {O.~S.}\ \bibnamefont
  {Magana-Loaiza}}, \bibinfo {author} {\bibfnamefont {I.~De}\ \bibnamefont
  {Leon}}, \bibinfo {author} {\bibfnamefont {M.}~\bibnamefont {Mirhosseini}},
  \bibinfo {author} {\bibfnamefont {R.}~\bibnamefont {Fickler}}, \bibinfo
  {author} {\bibfnamefont {A.}~\bibnamefont {Safari}}, \bibinfo {author}
  {\bibfnamefont {U.}~\bibnamefont {Mick}}, \bibinfo {author} {\bibfnamefont
  {B.}~\bibnamefont {McIntyre}}, \bibinfo {author} {\bibfnamefont
  {P.}~\bibnamefont {Banzer}}, \bibinfo {author} {\bibfnamefont
  {B.}~\bibnamefont {Rodenburg}}, \bibinfo {author} {\bibfnamefont
  {G.}~\bibnamefont {Leuchs}}, \ and\ \bibinfo {author} {\bibfnamefont {R.~W.}\
  \bibnamefont {Boyd}},\ }\bibfield  {title} {\enquote {\bibinfo {title}
  {Exotic looped trajectories of photons in three-slit interference},}\
  }\href@noop {} {\bibfield  {journal} {\bibinfo  {journal} {Nat. Comms.}\
  }\textbf {\bibinfo {volume} {7}},\ \bibinfo {pages} {13987} (\bibinfo {year}
  {2016})}\BibitemShut {NoStop}%
\bibitem [{\citenamefont {Rengaraj}\ \emph {et~al.}(2018)\citenamefont
  {Rengaraj}, \citenamefont {Prathwiraj}, \citenamefont {Sahoo}, \citenamefont
  {Somashekhar},\ and\ \citenamefont {Sinha}}]{Sinha2018}%
  \BibitemOpen
  \bibfield  {author} {\bibinfo {author} {\bibfnamefont {G.}~\bibnamefont
  {Rengaraj}}, \bibinfo {author} {\bibfnamefont {U.}~\bibnamefont
  {Prathwiraj}}, \bibinfo {author} {\bibfnamefont {S.~N.}\ \bibnamefont
  {Sahoo}}, \bibinfo {author} {\bibfnamefont {R.}~\bibnamefont {Somashekhar}},
  \ and\ \bibinfo {author} {\bibfnamefont {U.}~\bibnamefont {Sinha}},\
  }\bibfield  {title} {\enquote {\bibinfo {title} {Measuring the deviation from
  the superposition principle in interference experiments},}\ }\href@noop {}
  {\bibfield  {journal} {\bibinfo  {journal} {New J. Phys.}\ }\textbf {\bibinfo
  {volume} {20}},\ \bibinfo {pages} {063049} (\bibinfo {year}
  {2018})}\BibitemShut {NoStop}%
\bibitem [{\citenamefont {{\.Z}yczkowski}(2008)}]{zyczkowski2008quartic}%
  \BibitemOpen
  \bibfield  {author} {\bibinfo {author} {\bibfnamefont {Karol}\ \bibnamefont
  {{\.Z}yczkowski}},\ }\bibfield  {title} {\enquote {\bibinfo {title} {Quartic
  quantum theory: an extension of the standard quantum mechanics},}\
  }\href@noop {} {\bibfield  {journal} {\bibinfo  {journal} {Journal of Physics
  A: Mathematical and Theoretical}\ }\textbf {\bibinfo {volume} {41}},\
  \bibinfo {pages} {355302} (\bibinfo {year} {2008})}\BibitemShut {NoStop}%
\bibitem [{\citenamefont {Daki{\'c}}\ \emph {et~al.}(2014)\citenamefont
  {Daki{\'c}}, \citenamefont {Paterek},\ and\ \citenamefont
  {Brukner}}]{dakic2014density}%
  \BibitemOpen
  \bibfield  {author} {\bibinfo {author} {\bibfnamefont {B}~\bibnamefont
  {Daki{\'c}}}, \bibinfo {author} {\bibfnamefont {Tomasz}\ \bibnamefont
  {Paterek}}, \ and\ \bibinfo {author} {\bibfnamefont {{\v{C}}}~\bibnamefont
  {Brukner}},\ }\bibfield  {title} {\enquote {\bibinfo {title} {Density cubes
  and higher-order interference theories},}\ }\href@noop {} {\bibfield
  {journal} {\bibinfo  {journal} {New Journal of Physics}\ }\textbf {\bibinfo
  {volume} {16}},\ \bibinfo {pages} {023028} (\bibinfo {year}
  {2014})}\BibitemShut {NoStop}%
\bibitem [{\citenamefont {Lee}\ and\ \citenamefont
  {Selby}(2017)}]{lee2017higher}%
  \BibitemOpen
  \bibfield  {author} {\bibinfo {author} {\bibfnamefont {Ciar{\'a}n~M}\
  \bibnamefont {Lee}}\ and\ \bibinfo {author} {\bibfnamefont {John~H}\
  \bibnamefont {Selby}},\ }\bibfield  {title} {\enquote {\bibinfo {title}
  {Higher-order interference in extensions of quantum theory},}\ }\href@noop {}
  {\bibfield  {journal} {\bibinfo  {journal} {Foundations of Physics}\ }\textbf
  {\bibinfo {volume} {47}},\ \bibinfo {pages} {89--112} (\bibinfo {year}
  {2017})}\BibitemShut {NoStop}%
\bibitem [{\citenamefont {Kauten}\ \emph {et~al.}(2014)\citenamefont {Kauten},
  \citenamefont {Pressl}, \citenamefont {Kaufmann},\ and\ \citenamefont
  {Weihs}}]{Kauten2014}%
  \BibitemOpen
  \bibfield  {author} {\bibinfo {author} {\bibfnamefont {T.}~\bibnamefont
  {Kauten}}, \bibinfo {author} {\bibfnamefont {B.}~\bibnamefont {Pressl}},
  \bibinfo {author} {\bibfnamefont {T.}~\bibnamefont {Kaufmann}}, \ and\
  \bibinfo {author} {\bibfnamefont {G.}~\bibnamefont {Weihs}},\ }\bibfield
  {title} {\enquote {\bibinfo {title} {Measurement and modeling of the
  nonlinearity of photovoltaic and geiger-mode photodiodes},}\ }\href@noop {}
  {\bibfield  {journal} {\bibinfo  {journal} {Rev. Sci. Instrum.}\ }\textbf
  {\bibinfo {volume} {85}},\ \bibinfo {pages} {063102} (\bibinfo {year}
  {2014})}\BibitemShut {NoStop}%
\bibitem [{\citenamefont {Lee}\ \emph {et~al.}(2020)\citenamefont {Lee},
  \citenamefont {Zhuo}, \citenamefont {Couteau}, \citenamefont {Wilkowski},\
  and\ \citenamefont {Paterek}}]{leeAtomic}%
  \BibitemOpen
  \bibfield  {author} {\bibinfo {author} {\bibfnamefont {Kai~Sheng}\
  \bibnamefont {Lee}}, \bibinfo {author} {\bibfnamefont {Zhao}\ \bibnamefont
  {Zhuo}}, \bibinfo {author} {\bibfnamefont {Christophe}\ \bibnamefont
  {Couteau}}, \bibinfo {author} {\bibfnamefont {David}\ \bibnamefont
  {Wilkowski}}, \ and\ \bibinfo {author} {\bibfnamefont {Tomasz}\ \bibnamefont
  {Paterek}},\ }\bibfield  {title} {\enquote {\bibinfo {title} {Atomic test of
  higher-order interference},}\ }\href {\doibase 10.1103/PhysRevA.101.052111}
  {\bibfield  {journal} {\bibinfo  {journal} {Phys. Rev. A}\ }\textbf {\bibinfo
  {volume} {101}},\ \bibinfo {pages} {052111} (\bibinfo {year}
  {2020})}\BibitemShut {NoStop}%
\bibitem [{\citenamefont {Calafell}\ \emph {et~al.}(2021)\citenamefont
  {Calafell}, \citenamefont {Rozema}, \citenamefont {Iranzo}, \citenamefont
  {Trenti}, \citenamefont {Jenke}, \citenamefont {Cox}, \citenamefont {Kumar},
  \citenamefont {Bieliaiev}, \citenamefont {Nanot}, \citenamefont {Peng} \emph
  {et~al.}}]{calafell2021giant}%
  \BibitemOpen
  \bibfield  {author} {\bibinfo {author} {\bibfnamefont {Irati~Alonso}\
  \bibnamefont {Calafell}}, \bibinfo {author} {\bibfnamefont {Lee~A}\
  \bibnamefont {Rozema}}, \bibinfo {author} {\bibfnamefont {David~Alcaraz}\
  \bibnamefont {Iranzo}}, \bibinfo {author} {\bibfnamefont {Alessandro}\
  \bibnamefont {Trenti}}, \bibinfo {author} {\bibfnamefont {Philipp~K}\
  \bibnamefont {Jenke}}, \bibinfo {author} {\bibfnamefont {Joel~D}\
  \bibnamefont {Cox}}, \bibinfo {author} {\bibfnamefont {Avinash}\ \bibnamefont
  {Kumar}}, \bibinfo {author} {\bibfnamefont {Hlib}\ \bibnamefont {Bieliaiev}},
  \bibinfo {author} {\bibfnamefont {S{\'e}bastien}\ \bibnamefont {Nanot}},
  \bibinfo {author} {\bibfnamefont {Cheng}\ \bibnamefont {Peng}},  \emph
  {et~al.},\ }\bibfield  {title} {\enquote {\bibinfo {title} {Giant enhancement
  of third-harmonic generation in graphene--metal heterostructures},}\
  }\href@noop {} {\bibfield  {journal} {\bibinfo  {journal} {Nature
  Nanotechnology}\ }\textbf {\bibinfo {volume} {16}},\ \bibinfo {pages}
  {318--324} (\bibinfo {year} {2021})}\BibitemShut {NoStop}%
\bibitem [{\citenamefont {Sakurai}(1994)}]{Sakurai}%
  \BibitemOpen
  \bibfield  {author} {\bibinfo {author} {\bibfnamefont {Jun~John}\
  \bibnamefont {Sakurai}},\ }\href@noop {} {\emph {\bibinfo {title} {{Modern
  quantum mechanics; rev. ed.}}}}\ (\bibinfo  {publisher} {Addison-Wesley},\
  \bibinfo {address} {Reading, MA},\ \bibinfo {year} {1994})\BibitemShut
  {NoStop}%
\bibitem [{\citenamefont {Namdar}\ \emph {et~al.}(2021)\citenamefont {Namdar},
  \citenamefont {Jenke}, \citenamefont {Alonso~Calafell}, \citenamefont
  {Trenti}, \citenamefont {Radonjic}, \citenamefont {Dakic}, \citenamefont
  {Walther},\ and\ \citenamefont {Rozema}}]{hoiDATA}%
  \BibitemOpen
  \bibfield  {author} {\bibinfo {author} {\bibfnamefont {Peter}\ \bibnamefont
  {Namdar}}, \bibinfo {author} {\bibfnamefont {Philipp~K.}\ \bibnamefont
  {Jenke}}, \bibinfo {author} {\bibfnamefont {Irati}\ \bibnamefont
  {Alonso~Calafell}}, \bibinfo {author} {\bibfnamefont {Alessandro}\
  \bibnamefont {Trenti}}, \bibinfo {author} {\bibfnamefont {Milan}\
  \bibnamefont {Radonjic}}, \bibinfo {author} {\bibfnamefont {Borivoje}\
  \bibnamefont {Dakic}}, \bibinfo {author} {\bibfnamefont {Philip}\
  \bibnamefont {Walther}}, \ and\ \bibinfo {author} {\bibfnamefont {Lee~A.}\
  \bibnamefont {Rozema}},\ }\href {\doibase 10.5281/zenodo.5775886} {\enquote
  {\bibinfo {title} {{Raw data for the manuscript "Experimental Higher- Order
  Interference in a Nonlinear Triple Slit"}},}\ } (\bibinfo {year}
  {2021})\BibitemShut {NoStop}%
\bibitem [{\citenamefont {Boyd}(2020)}]{boyd2020nonlinear}%
  \BibitemOpen
  \bibfield  {author} {\bibinfo {author} {\bibfnamefont {Robert~W}\
  \bibnamefont {Boyd}},\ }\href@noop {} {\emph {\bibinfo {title} {Nonlinear
  optics}}}\ (\bibinfo  {publisher} {Academic press},\ \bibinfo {year}
  {2020})\BibitemShut {NoStop}%
\bibitem [{\citenamefont {{AS Photonics}}(2020)}]{SNLO}%
  \BibitemOpen
  \bibfield  {author} {\bibinfo {author} {\bibnamefont {{AS Photonics}}},\
  }\href@noop {} {\enquote {\bibinfo {title} {{SNLO} v.75},}\ }\bibinfo
  {howpublished} {\url{https://as-photonics.com/}, 08/10/2020} (\bibinfo {year}
  {2020})\BibitemShut {NoStop}%
\end{thebibliography}%
\clearpage

\section*{Appendix}

\subsection*{Classical Description and Fitting to Data}

In classical nonlinear optics, the central object is the nonlinear polarization, which describes the coupling of the incident light fields to a nonlinear medium, which gives rise to re-emission of light at potentially different frequencies.
For example, two fields at frequencies $\omega_1$ and $\omega_2$ can excite a nonlinear polarization associated with either sum-frequency generation $P(\omega_1+\omega_2)=2\epsilon_o\chi^{(2)}E_1E_2$ or difference frequency generation $P(\omega_1-\omega_2)=2\epsilon_o\chi^{(2)}E_1E_2^*$, where $E_1$ and $E_2$ are the incident fields.
Notice that any phase between fields $E_1$ and $E_2$ becomes a global phase on the polarization. Hence, processes with only two input fields are independent of this phase.
If, however, three fields are incident, as is the case for our experiment, the relative phase between the three beams is no-longer global. Hence, the phase can, in general, affect the nonlinear generation.
To see this, consider three fields $E_1$, $E_2$, and $E_3$, the first two at a frequency $\omega$, and the third at $2\omega$, that are incident in a $\chi^{(2)}$ nonlinear medium that is  phase-matched for both SFG between $E_1$ and $E_2$, and DFG between both $E_1$ and $E_3$, and $E_2$ and $E_3$.
In this case, the polarization will be
\begin{equation}\label{eq:phase}
    2\epsilon_o\chi^{(2)}\left(E_3E_1^*e^{i(\phi_3-\phi_1)}+E_3E_2^*e^{i(\phi_3-\phi_2)}+E_1E_2e^{i(\phi_2+\phi_1)}\right).
\end{equation}
Here we have explicitly separated the phases $\phi_1$, $\phi_2$, and $\phi_3$ from their respective fields to stress the phase dependence.

To solve for the generated fields with three incident beams we use the so-called Coupled Wave Equations, as described in Ref.~\cite{boyd2020nonlinear}:
\begin{eqnarray}
\frac{dE_{1}}{dz}=\frac{2id_\mathrm{eff}\omega_1^2}{k_1 c^2} E_3 E_2^*e^{-i\Delta kz}\\
\frac{dE_{2}}{dz}=\frac{2id_\mathrm{eff}\omega_2^2}{k_2 c^2} E_3 E_1^*e^{-i\Delta kz}\\
\frac{dE_{3}}{dz}=\frac{2id_\mathrm{eff}\omega_3^2}{k_3 c^2} E_1 E_2e^{-i\Delta kz}.
\end{eqnarray}
In these equations $\Delta k$ is the phase mismatch, which we will set to $0$ for our simple model, $d_\mathrm{eff}$ is the nonlinear strength of the crystal, and $k_i$ is the wave vector of each field.
Setting $g=\frac{2 d_\mathrm{eff}\omega_1}{c}$, and $\omega_\mathrm{1} = \omega_\mathrm{2} = \frac{1}{2}\omega_\mathrm{3}$, we arrive Eqs. \ref{eq:scde1}-\ref{eq:scde3} of the main text, which can be solved numerically.

Then to compute a value for the Sorkin parameter $\kappa$, we simply solve these equations, under different initial conditions to compute each term of $\kappa$ in Eq. \ref{eq:sorkin_parameter}. 
For example, to compute $P_{12}$, we use the initial conditions $E_1(z=0)=\sqrt{\mathcal{P}_1}$, $E_2(z=0)=\sqrt{\mathcal{P}_2}$, and $E_3(z=0)=0$, where $\mathcal{P}_i$ is the power input into mode $i$.
Then we evaluate the field in mode 3 (the pump mode) at $z=1$ mm (the length of the crystal) to compute the optical power in mode 3 after exiting the crystal.  
While this is sufficient to show a non-zero value of $\kappa$, to model our experiment we include additional experiment factors: (1) the changing efficiency of the nonlinear processes during the z-scan and the different relative efficiencies of the three nonlinear processes, and (2) the optical phase induced during the z-scan.

\begin{figure}
    \centering
    \includegraphics[width=\columnwidth]{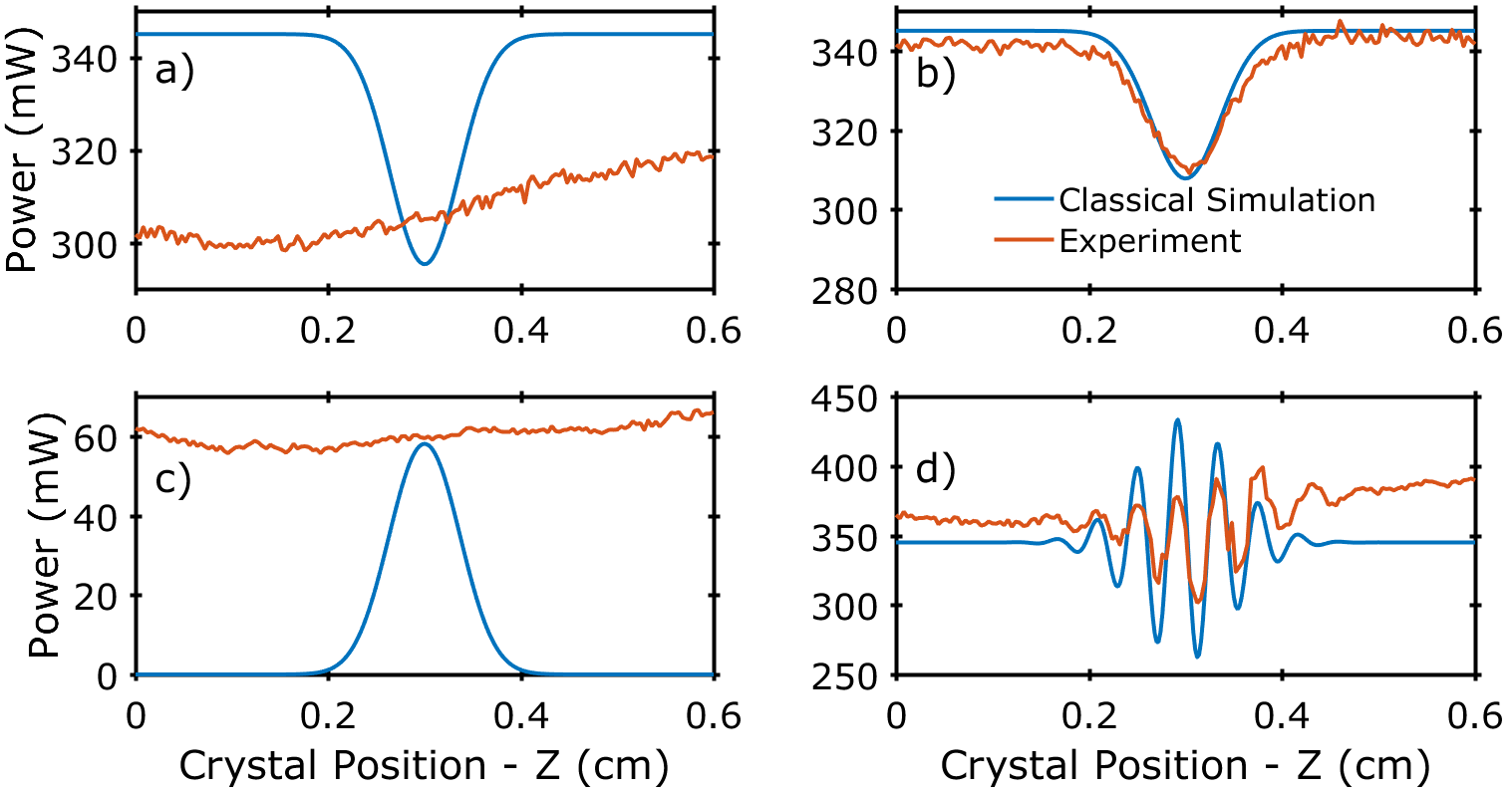}
    \caption{Individual Sorkin terms versus position of the crystal. The orange curves show the power that is measured in the pump mode 3 after the crystal. For the data in panels a) ($\mathcal{P}_{13}$) and b) ($\mathcal{P}_{23}$) the pump beam undergoes difference-frequency generation with the signal and idler, respectively.  The dip decrease in power corresponds to difference-frequency power being generated in the third mode. c) shows sum-frequency generation between the signal and idler into the pump mode ($\mathcal{P}_{12}$).  Over the range the crystal is scanned here, the power is more or less constant.
    d) For these data, all three beams are open and the three process all occur simultaneously $\mathcal{P}_{123}$. The oscillations are induced by a phase discussed in the main text.
    In all panels, the blue curves are the results of the simulations used to fit to the data.  For simplicity, we fix the width of the interaction region in or simulation to the narrowest process shown in panel b). This is why the simulations presented in panels a) and c) only match the measured values at the focus.
    }
    \label{fig:DFG_SFG}
\end{figure}

\subsubsection{Varying Efficiencies}
To model varying efficiencies we first use a different coupling constant for each differential equation:
\begin{eqnarray}
\frac{dE_{1}}{dz}=i g_1(Z) E_3 E_2^*\\
\frac{dE_{2}}{dz}=i g_2(Z) E_3 E_1^*\\
\frac{dE_{3}}{dz}=i 2 g_3(Z) E_1 E_2,
\end{eqnarray}
and we also make them depend on the crystal position $Z$.
We use different coupling constants to model the fact that, say, beam 1 and 2 may spatially and temporally overlap differently than beams 2 and 3.
Then we numerically solve these equations for  different values of $Z$.
In particular, we assume a Gaussian form for $g_i(Z)$:
\begin{equation}\label{eq:gauss}
    g_i(Z) = \eta_i ge^{-\frac{2Z^2}{\Delta^2}},
\end{equation}
where $\Delta$ characterizes the interaction range, $g$ is coupling constant defined below Eq. \ref{eq:scde1}-\ref{eq:scde3}, and $\eta_i$ is a fitting parameter we use to account for experimental imperfections such as beam overlap, and walk-off in the crystal.

The goal of this is to model the different interactions between the three beams as the crystal is translated through the focus.
Experimentally, this effect is evident in different data sets. For example, the data presented in Fig. \ref{fig:processEff} shows that as the crystal is moved through the intersection points of the beams the different processes respond differently.
We believe that this is caused by slightly different intersection angles between the three beams, and by the fact that the crystal is not perfectly perpendicular to the pump beam or the direction the crystal is moved.

To fit to our data, we observe the power in the pump mode $3$ when all pairs of beams are turned on and the crystal is scanned through the focus. In particular we measure the Z-dependence of $\mathcal{P}_{13}$ (Fig. \ref{fig:DFG_SFG}a), 
$\mathcal{P}_{23}$ (Fig. \ref{fig:DFG_SFG}b), and $\mathcal{P}_{12}$ (Fig. \ref{fig:DFG_SFG}c).
For these sets of measurements, the input power in each beam, is $870$ mW in the signal (mode 1), $600$ mW in the idler (mode 2), and $345$ mW in the pump (mode 3).
For our simulation we then convert these average powers to fields, assuming Gaussian pulses with a width of $\tau=140$ fs, repetition rate of $R=76$ MHz, and beam diameter at the focus of $w=26~\mu$m.
Then the field is
\begin{equation}
    E=\sqrt{ \left(\frac{\mathrm{ln}2}{\pi}\right)^{3/2}\frac{16}{R\tau w^2\epsilon_0 c} \mathcal{P}},
\end{equation}
where $\epsilon_0$ is the permittivity of free space and $c$ is the speed of light \cite{calafell2021giant}.
In our simulation, we set the initial conditions to match the experimental configuration (one beam blocked and the other two open), and then adjust $\eta_i$ to match the change in power observed throughout the z-scan.
The result of these fits is shown in Fig. \ref{fig:DFG_SFG}a-c.
For simplicity we assume the same the interaction length for all three processes, $\Delta=0.05$ cm, which is given by narrowest process, which is DFG between the idler (Fig. \ref{fig:DFG_SFG}b).
We set $\eta_1=\eta_2=0.15$ and $\eta_3=0.05$, and use $d_\mathrm{eff}=0.749$ pm/V (given in \cite{SNLO} for LBO) to evaluate $g$.
Additionally, we scale all three values of $\eta$ by an additional factor of $0.3$ when all three beams are open---i.e. when measuring $\mathcal{P}_{123}$---to account for the reduced `mutual overlap' between all three beams.

\subsubsection{Phase Induced During Z-Scan}

\begin{figure}
    \centering
    \includegraphics[width=\columnwidth]{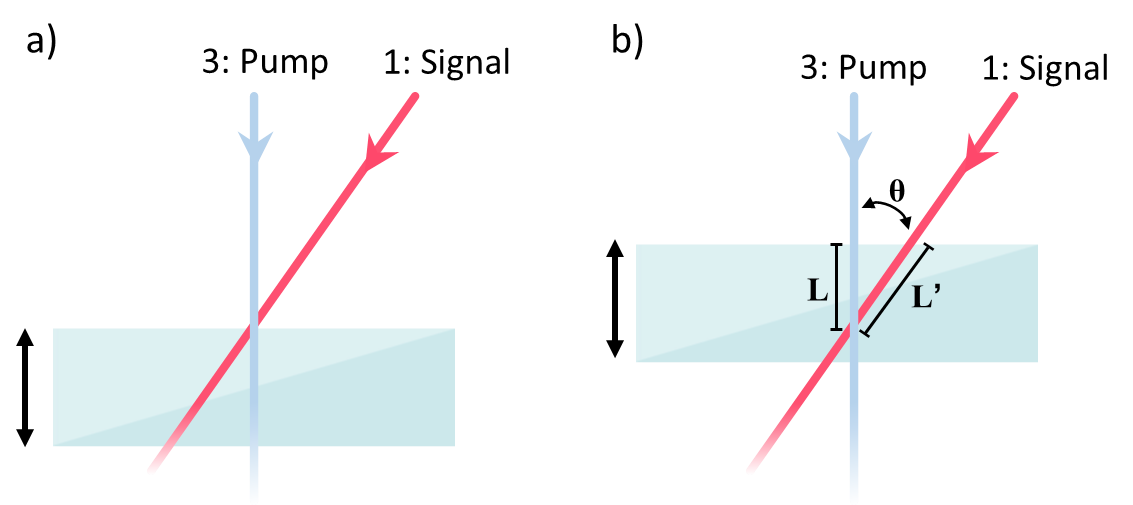}
    \caption{Phase acquired during Z-scan. a) At the start of the Z-scan of the crystal the pump and signal beams intersect in front of the crystal. We set this to be a zero relative phase between the two beams.
    b) As the crystal is translated through the intersection point, the optical path length for the pump beam $L$ changes slower than the optical path of the signal beam $L'$. Thus as the crystal is translated through the intersection point the relative phase is also scanned.}
    \label{fig:xtalPhase}
\end{figure}

To acquire our data, we translate the crystal along $Z$, through the focus.
In addition to modulating the strength of the nonlinear interaction, this induces a phase between the pump beam and the signal and idler beams.
The phase is symmetric, i.e. the induced phase between the pump and signal is the same as the induced phase between the pump and idler.
This effect, for the signal and pump beams, is illustrated in Fig. \ref{fig:xtalPhase}.
Panel a shows the situation when the two beams intersect before the crystal. We define this as our $0$ phase reference.
As the crystal is moved through the focus the optical path the pump experiences changes differently from that the signal experiences (panel b).
A straightforward calculation shows that the pump picks up phase as $\phi_\mathrm{3}=\frac{2\pi n}{\lambda_3} Z$, where $n=1.611$ is the refractive index of our LBO crystal~\cite{SNLO} (since the crystal is phase matched for degenerate SFG this is the same for all three beams) and $\lambda_3=400$ nm.
Similarly, the signal and idler acquire phase as $\phi_\mathrm{1,2}=\frac{2\pi n}{\lambda_{1,2}\cos{\theta}}Z$, where $\lambda_{1,2}=800$ nm.

\subsection*{Quantum Mechanical Description}

To express our experiment within quantum mechanics, we start with the Hamiltonian $\Hat{H}$ describing three modes interacting in a $\chi^{(2)}$-nonlinear medium: 
\begin{equation}
\begin{split}
\hat{H}=~&\hbar\omega \hat{a}^\dagger_1\hat{a}_1 + \hbar\omega \hat{a}^\dagger_2\hat{a}_2 + 2\hbar\omega \hat{a}^\dagger_3\hat{a}_3 \\&+ i\hbar\chi^{(2)} (\hat{a}_1\hat{a}_2\hat{a}_3^\dagger - \hat{a}_1^\dagger\hat{a}_2^\dagger\hat{a}_3).
\label{eq:hamiltonian}
\end{split}
\end{equation} 
Here, $\hat{a}^\dagger_i$ and $\hat{a}_i$ are the ladder operators acting on mode $i$.
As is in the main text, mode 1 describes the signal beam at frequency $\omega$, mode 2 the idler beam at $\omega$, and mode 3 the pump beam at $2\omega$.
Since we measure photons in mode 3 after the interaction, we need to evolve the number operator associated with mode 3 as
\begin{equation}
    \hat{n}_{3,f}= e^{i\Hat{H}\tau/\hbar}\hat{a}^\dagger_3\hat{a}_3e^{-i\Hat{H}\tau/\hbar}.
    \label{eq:tobeexpanded}
\end{equation}
Here, $\tau$, the interaction time, is proportional to the crystal length.
We then apply a Baker-Campbell-Haussdorff expansion \cite{Sakurai} up to 6th order to Eq.  \ref{eq:tobeexpanded}.
Given the number of terms in the expansion we do not include the result here, but it is available in our Mathematica script uploaded to \cite{hoiDATA}.
Since we will only consider coherent input states $\ket{\alpha_i}$, we can do the following substitution into our expanded form of $\hat{n}_{3,f}$:
\begin{equation}
\begin{split}
    \hat{a}_i\ket{\alpha_i} &\rightarrow \sqrt{n_i}~e^{i\phi_i} \\ \bra{\alpha_i}\hat{a}_i^\dagger &\rightarrow \sqrt{n_i}~e^{-i\phi_i},
\end{split}
\end{equation}
where the phase factors $\phi_i$ represent the phase acquired during the Z-scan, and are defined in the previous section.
We are now interested in the average photon number in mode 3 after the interaction.
This is computed by taking the expectation value of $\hat{n}_{3,f}$ for coherent inputs as:
\begin{equation}\label{eq:expect}
\begin{split}
     \bra{\alpha_1, \alpha_2, \alpha_3}~\hat{n}_{3,f}\ket{\alpha_1, \alpha_2, \alpha_3} &=\\ f(\tau,\kappa(z), n_1, n_2,& n_3, \phi_1(z), \phi_2(z), \phi_3(z)).
\end{split}
\end{equation}
From this equation we compute the outcome photon number for all experimental configurations by inserting the specific states for $\alpha_i$.
For example, if all three beams are open, the outcome photon number in mode 3 is simply
\begin{equation}
n_{123} = \bra{\alpha_1, \alpha_2, \alpha_3}\hat{n}_{3,f} \ket{\alpha_1, \alpha_2, \alpha_3}.
\end{equation}
If, e.g., the pump beam is blocked the output photon number is
\begin{equation}
n_{12} = \bra{\alpha_1, \alpha_2, 0}\hat{n}_{3,f} \ket{\alpha_1, \alpha_2, 0},
\label{eq:two_beams}
\end{equation}
and so forth. 
As in the classical simulations, the relative phase cancels out for all configurations except $n_{123}$ and has the same form as stated there.
From the output photon numbers we then compute the individual terms of Eq. \ref{eq:sorkin_parameter} by normalizing by the input photon number.
For example, $P_{12}=n_{12}/\left(|\alpha_1|^2+|\alpha_2|^2+|\alpha_3|^2\right)$, and so on.

\subsubsection{Fitting to the Nonlinear Strength}
To fit to our experimental data, we first estimate the product of the nonlinearity and the interaction time, which we define as $\Gamma = \tau\chi^{(2)}$.
To do this, we examine two interacting beams in the crystal.
In particular, we examine the SFG process by sending the signal and idler beams into our nonlinear crystal and measuring the sum-frequency light generated in the pump beam, as in the configuration $n_{12}$. 
We then vary the input powers and measure the output power.
These  powers are then all converted into photon numbers $n_{1}$, $n_{2}$ and ${n_{3, f}}$.

\begin{figure}
    \centering
    \includegraphics[width=1.1\linewidth]{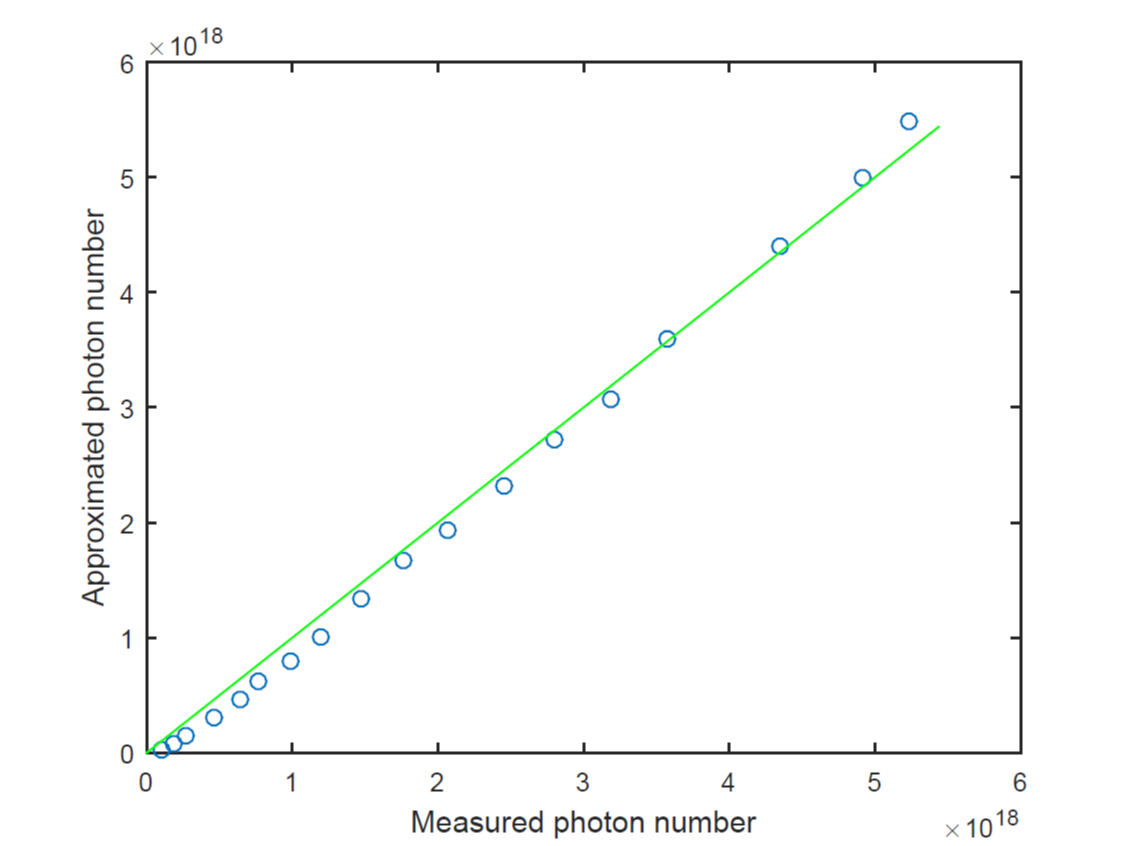}
    \caption{Extraction of the nonlinear interaction $\Gamma$ from measured data. The blue points are a plot of the approximated photon numbers versus the measured photon numbers. Note we measure the average power, and then compute the photon number. Error bars are not plotted, since the uncertainty is on the order of the size of the circles. For an ideal approximation, this procedure should result in a line with a slope of 1 (the green line).}
    \label{fig:kappa_fit}
\end{figure}

Under these conditions, the only unknown from Eq. \ref{eq:two_beams}, is $\Gamma$. 
We then compute the residuals between the measured data points and the predictions of Eq. \ref{eq:two_beams}. By minimizing the residuals over $\Gamma$ we obtain $\Gamma = 1.05\times10^{-6}$. In Fig. \ref{fig:kappa_fit} our approximation is visualized by plotting the predictions of Eq. \ref{eq:two_beams} versus the measured  power (blue points).
Ideally this results in a line with a slope of 1 (green line).
The close fit demonstrates that we have sufficient precision using the 6th order expansion.

Finally, to model the Z-scan of the crystal through the focus, we take $\Gamma$ to be a Gaussian function of $Z$, as in Eq. \ref{eq:gauss} of the previous section.
We then simply compute $\kappa$ by computing the individual terms of Eq. \ref{eq:sorkin_parameter} using Eq. \ref{eq:expect} for each crystal position Z where both $\Gamma$ and the phase are functions of $Z$.
The result, shown in the lower left insert in Fig. \ref{fig:main_result}, reproduces the qualitative features of our experiment.

\newpage
\onecolumngrid

\subsection{Additional Measurements}

\begin{figure*}[h]
    \centering
    \includegraphics[width=\linewidth]{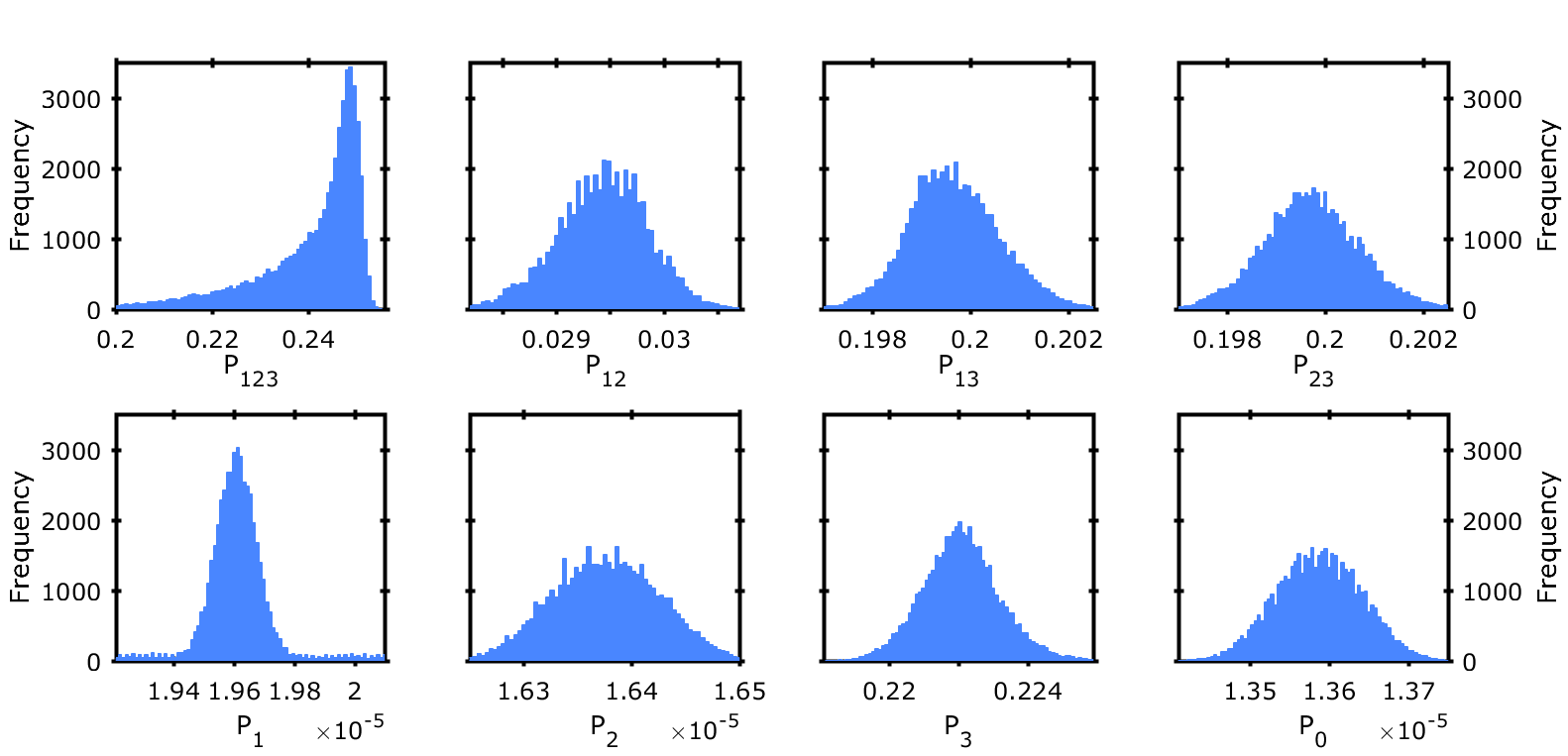}
    \caption{Histograms of the individual terms of the Sorkin-parameter. 
    The $x$-axis for these data is the normalized power, as described in the main text, for a given configuration of the experiment. For example, $P_{12}$ is the normalized power measured when beams 1 and 2 are opened, and beam 3 is blocked.
    These data were used to compute the final value of $\kappa=0.0334\pm0.0002$, the data presented in Fig. \ref{fig:sorkin_static}.}
    \label{fig:individual_histograms}
\end{figure*}

\begin{figure*}
    \centering
    \includegraphics[width=0.5\columnwidth]{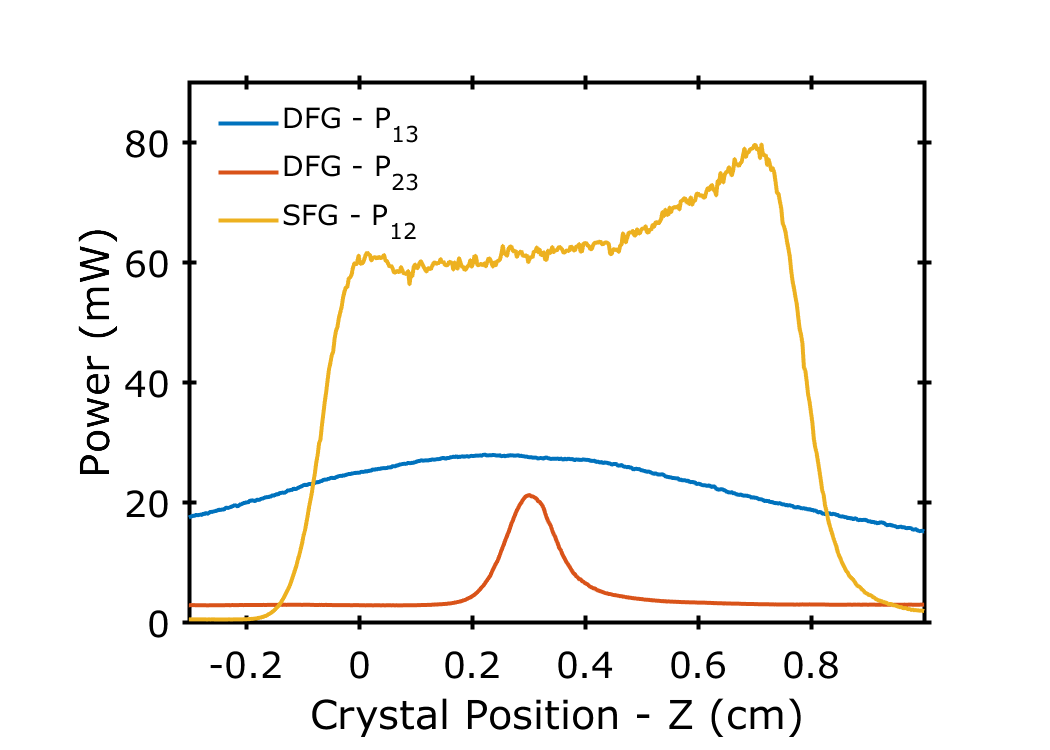}
    \caption{Generated power for each two-mode process.  For these data, one path is blocked, and a detector is placed after the crystal in the blocked path. 
    In other words, the detector is moved among the modes to ensure that the power generated by the nonlinear mixing is always measured.
    Then the power is recorded as the crystal is translated through the beams focus. 
    From the vastly different Z dependence of these data, one can observe that the different pairs of beams have significantly different overlap within the crystal.
    }
    \label{fig:processEff}
\end{figure*}

\begin{figure*}
    \centering
    \includegraphics[width=0.5\columnwidth]{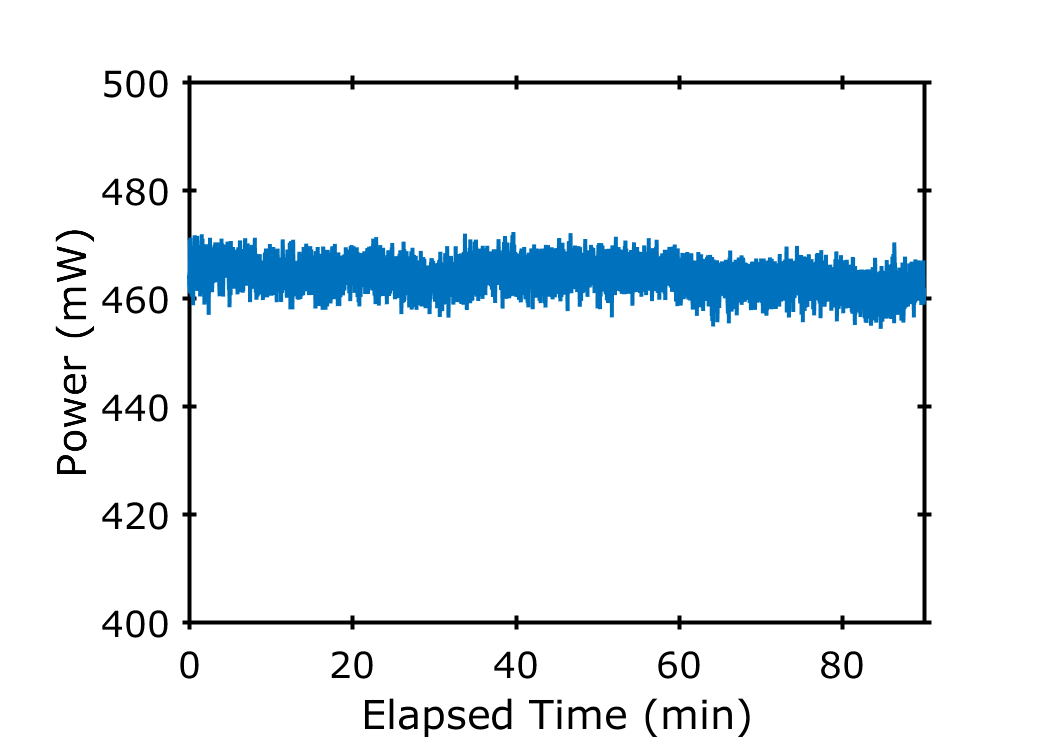}
    \caption{\textbf{Stability of $\mathcal{P}_{123}$.} The power detected in the pump beam when all three beams are present plotted versus time.  For these data, the crystal position was fixed and all three paths were left open. The phase in our experiment is stable for over 90 minutes.
    }
    \label{fig:stability}
\end{figure*}

\end{document}